\documentclass[reprint,aps,superscriptaddress]{revtex4-1}
\usepackage{amssymb}
\usepackage{amsmath}
\usepackage{graphicx}
\usepackage{mathdots}
\usepackage{xcolor}
\usepackage{xspace}
\usepackage[toc,page]{appendix}

\begin{document}

\title{Floquet-Sambe Bottleneck and Frequency-Selective Localization in a
Driven Synthetic Spin Chain}
\author{J. Cao}
\affiliation{College of Physics and Materials Science, Tianjin Normal
University, Tianjin 300387, China}
\author{K. L. Zhang}
\affiliation{School of Physics and Optoelectronic Engineering, Foshan
University, Foshan, 528225, China}
\author{R. Wang}
\email{wangr@tjnu.edu.cn}
\affiliation{College of Physics and Materials Science, Tianjin Normal
University, Tianjin 300387, China}
\author{X. Z. Zhang}
\email{zhangxz@tjnu.edu.cn}
\affiliation{College of Physics and Materials Science, Tianjin Normal
University, Tianjin 300387, China}

\begin{abstract}
We study a finite Floquet chain in which a uniform nearest-neighbor hopping
coexists with a periodically rotating, \textrm{SU(2)}-dictated spin-assisted
hopping profile. The resulting coupling is spatially inhomogeneous---weakest
at the chain boundaries and strongest in the bulk---and produces a
frequency-dependent Floquet-Sambe bottleneck. In the closed system, the
mean inverse participation ratio (\textrm{MIPR}) of the Floquet eigenstates
exhibits a striking nonmonotonic dependence on the driving frequency $\omega
$: the states remain extended at both low and high frequencies, but become
maximally localized at an intermediate frequency. We demonstrate that this
localization maximum occurs at $\omega _{\mathrm{peak}}\sim \mu _{-s}=\sqrt{%
2s}$, a scale controlled by the first boundary bottleneck. To connect these
spectral properties to measurable transport, we construct an open-system
Floquet-Sambe Green-function inverse participation ratio from the spatial
density of the injected scattering state. This open-system diagnostic
recovers the same nonmonotonic localization trend as its closed-system
counterpart, with the peak shifted to higher frequencies by the static
bandwidth and the lead self-energy. These findings establish the driven
synthetic spin chain as a directly realizable, frequency-tunable platform
for coherent information storage and retrieval, rooted in the interplay of
Floquet-Sambe virtual channels, boundary-controlled localization, and
frequency-selective transport in emerging multi-level superconducting
circuit architectures.
\end{abstract}

\maketitle

\section{Introduction}

Coherent control via periodic driving---termed Floquet engineering \cite%
{Takashi}---has become a central paradigm in non-equilibrium quantum physics
\cite{AdvP2015,RMP2017,NRP2020,ARCMP2019,ARCMP2020}. Building on early
theoretical frameworks by Shirley and Sambe \cite{Shirley,Sambe}, it has
undergone a renaissance driven by rapid advances in laser technology and
ultrafast spectroscopy, together with the discovery of quantum materials
with exotic, tunable properties. Periodically driven systems, governed by $%
H(t+T)=H(t)$, admit a Floquet-Sambe representation in which the Hilbert
space is extended to include Fourier sectors. The resulting static
eigenvalue problem yields quasi-energies $\varepsilon _{n}\in \lbrack
-\omega /2,\omega /2)$ and Floquet modes that encode both the stroboscopic
dynamics and the inter-band dressing induced by the drive. Floquet
engineering thus offers a systematic route to realizing effective
Hamiltonians inaccessible in equilibrium, giving rise to phenomena ranging
from anomalous topological phases to dynamical localization and
light-induced superconductivity. These ideas have been explored across
diverse platforms, including ultracold atoms \cite%
{RMP2017,PRL2016,PRX2014,NP2019,NP2020,PRX2020,PRB2022}, photonic lattices
\cite{Nat2013,PRL2019}, superconducting qubits \cite{NP2017}, and graphene
\cite{NP2020-2}.

In lattice systems, the Floquet-Sambe picture is especially illuminating:
different Fourier components of the wave-function appear as photon sectors
separated by the drive frequency $\omega $, and hybridization among these
sectors generates effective static potentials and renormalized hopping
amplitudes. A particularly structured setting is provided by spin-assisted
tight-binding chains, in which the hopping amplitudes are set by \textrm{%
SU(2)} angular-momentum matrix elements and are therefore spatially
inhomogeneous---weakest at the two boundaries and strongest in the bulk.
This algebraic structure is naturally connected to the concept of synthetic
dimensions, where a set of internal states \cite{RMP2017} or energy levels
of a superconducting qudit \cite{PRX2025} is reinterpreted as a lattice
coordinate, and couplings between neighboring internal states play the role
of hopping amplitudes. Periodic driving of such systems can generate
synthetic gauge fields, artificial magnetic fluxes, and programmable
high-dimensional spin dynamics \cite{PRAP2018,PRA2019,PRL2025,NP2023}.

A canonical spin-assisted chain is the Christandl chain \cite{Christdl},
whose \textrm{SU(2)} hopping profile endows it with perfect quantum-state
transfer at a fixed, algebraically determined time. Its transport properties
are entirely fixed by the static algebraic structure, leaving no room for
in-situ control. These developments---Floquet engineering of
synthetic-dimension lattices on one hand, and the rigid algebraic structure
of the Christandl chain on the other---motivate the following question: Can
a monochromatic drive convert this algebraically engineered chain into a
tunable Floquet lattice whose localization and transport properties are
controlled by the driving frequency?

In this work, we study a finite Floquet chain in which uniform
nearest-neighbor hopping is combined with a periodically rotating,
spin-assisted hopping profile fixed by the \textrm{SU(2) }matrix elements of
the Christandl model. The coupling is spatially inhomogeneous and produces a
frequency-dependent Floquet-Sambe bottleneck at the chain boundaries. We
find that the mean inverse participation ratio (\textrm{MIPR})---a standard
measure of wavefunction localization---of the Floquet eigenstates exhibits a
striking nonmonotonic dependence on the driving frequency $\omega $. In the
low-frequency limit ($\omega \rightarrow 0$), all Floquet sectors are nearly
degenerate and the eigenstates remain extended. As $\omega $ increases
toward an intermediate scale, inter-sector hybridization is controlled by
the weakest link at the chain boundary, causing the states to localize
maximally at $\omega _{\mathrm{peak}}\sim \sqrt{2s}$, a scale set by the
first boundary bottleneck. In the high-frequency regime, the Magnus
expansion generates an effective Stark term $\propto S_{z}/\omega $; as $%
\omega $ increases further, this Stark field weakens, the Wannier-Stark
ladder collapses, and the eigenstates become extended again. To connect
these closed-system spectral properties to measurable transport, we
introduce an open-system Floquet-Sambe Green-function inverse participation
ratio (\textrm{IPR}) constructed from the spatial density of the injected
scattering state. This diagnostic recovers the same nonmonotonic
localization trend, with the peak shifted to higher frequencies by the
static bandwidth and the lead self-energy, establishing consistency between
the closed- and open-system perspectives. Together, these results reveal a
direct and tunable connection between Floquet-Sambe virtual channels,
boundary-controlled localization, and frequency-selective transport in a
driven synthetic spin chain.

The remainder of this paper is organized as follows. In Sec. \ref{Model
Hamiltonian}, we introduce the periodically driven spin-assisted
tight-binding model and its Floquet-Sambe representation, including the
structure of the quasi-energy spectrum. Section \ref{bottleneck} analyzes
the boundary-controlled Floquet-Sambe bottleneck and derives the
characteristic scale $\omega _{\text{\textrm{peak}}}\sim \sqrt{2s}$
governing the localization maximum. In Sec. \ref{SL Regime}, we discuss the
high-frequency effective Hamiltonian obtained from the Magnus expansion, the
emergence of a Stark-ladder description, and the unified picture connecting
the closed-system \textrm{MIPR} to the open-system Green-function \textrm{IPR%
}. Section \ref{Summary} summarizes our findings.

\section{Model and Floquet Formulation}

\label{Model Hamiltonian}

Perfect state transfer in engineered spin networks provides a minimal
protocol for coherent quantum communication without active control of
individual qubits during the transfer process. A central result is that a
nearest-neighbor chain with engineered couplings
\begin{equation}
J_{n}=J_{0}\sqrt{n(N-n)}  \label{Jn}
\end{equation}%
realizes mirror-symmetric state transfer across an $N$-site lattice at a
prescribed time. The algebraic origin of Eq. (\ref{Jn}) is the
representation theory of \textrm{SU(2)}. If the $N=2s+1$ lattice sites are
identified with the spin states $|s,j\rangle $, where $j=-s$, $-s+1$,\ldots
, $s$, the spin ladder operators generate nearest-neighbor hopping in the
site basis
\begin{equation}
S_{+}=\sum_{j=-s}^{s-1}\mu _{j}|j+1\rangle \langle j|,\text{ }%
S_{-}=\sum_{j=-s}^{s-1}\mu _{j}|j\rangle \langle j+1|,
\end{equation}%
with
\begin{equation}
\mu _{j}=\sqrt{s(s+1)-j(j+1)}.  \label{u_j}
\end{equation}%
Thus the inhomogeneous hopping profile is not an externally imposed random
structure, but follows from the matrix elements of a collective spin. This
observation underlies the Christandl type chain and explains why its static
dynamics can be mapped to spin precession generated by $%
S_{x}=(S_{+}+S_{-})/2 $.

We consider a finite open-boundary chain with Hamiltonian
\begin{equation}
H(t)=H_{0}+H_{s}(t),  \label{H_og}
\end{equation}%
where $H_{0}$ describes a uniform nearest-neighbor tight-binding chain,
\begin{equation}
H_{0}=-\kappa \sum_{j=-s}^{s-1}\left( |j\rangle \langle j+1|+\mathrm{H.c.}%
\right)
\end{equation}%
with the hopping amplitude $\kappa $. The time-dependent part models a
coherent spin drive,
\begin{equation}
H_{s}(t)=\sum_{j=-s}^{s-1}\left[ e^{-i\omega t}\mu _{j}\left\vert
j\right\rangle \left\langle j+1\right\vert +e^{i\omega t}\mu _{j}\left\vert
j+1\right\rangle \left\langle j\right\vert \right] .
\end{equation}%
Equivalently, Eq. (\ref{H_og}) becomes
\begin{equation}
H(t)=\sum_{j=-s}^{s-1}(-\kappa +e^{-i\omega t}\mu _{j})(|j\rangle \langle
j+1|)+\mathrm{H.c.}.  \label{H_1}
\end{equation}%
The ($2s+1$)-dimensional spin manifold is thus mapped onto a one-dimensional
chain spanned by basis $\{|j\rangle \}$. This mapping is physically
equivalent to constructing a synthetic dimension: the spin degree of freedom
acts as an extra discrete spatial coordinate, and $H_{s}(t)$ drives hopping
along this synthetic direction.

\subsection{Floquet-Sambe formulation}

For a finite driving frequency $\omega $, the Hamiltonian is periodic,
\begin{equation}
H(t+T)=H(t),\text{ }T=\frac{2\pi }{\omega }.
\end{equation}%
Floquet theory applies directly and allows one to rewrite the time-dependent
problem as a time-independent eigenvalue problem in an enlarged Hilbert
space. Solutions of the Schr\"{o}dinger equation take the form
\begin{equation}
|\psi _{j}(t)\rangle =e^{-i\varepsilon _{j}t}|\phi _{j}(t)\rangle ,
\label{phi_b}
\end{equation}%
where the Floquet modes satisfy $|\phi _{j}(t+T)\rangle =|\phi
_{j}(t)\rangle $. Substituting Eq. (\ref{phi_b}) yields
\begin{equation}
\left[ H(t)-i\partial _{t}\right] |\phi _{j}(t)\rangle =\varepsilon
_{j}|\phi _{j}(t)\rangle .  \label{eq_fs}
\end{equation}%
Eq. (\ref{eq_fs}) becomes a time-independent eigenvalue problem when
formulated in the Sambe space
\begin{equation}
\mathcal{S}=\mathcal{H}\otimes \mathcal{T},
\end{equation}%
where $\mathcal{T}$ is the space of $T$-periodic functions. The physical
picture is that Fourier components of the wavefunction correspond to
different photon sectors, so that $\varepsilon _{j}$ describes energies
modulo $\omega $, forming \textquotedblleft Floquet Brillouin
zones\textquotedblright\ analogous to quasi-momentum zones in spatial
crystals.

Expanding the Floquet modes and Hamiltonian in Fourier series,
\begin{equation}
|\phi _{j}(t)\rangle =\sum_{m=-\infty }^{\infty }e^{-im\omega t}|\phi
_{j}^{(m)}\rangle ,\text{ }H(t)=\sum_{n=-\infty }^{\infty }e^{-in\omega
t}H^{(n)},
\end{equation}%
and substituting into Eq. (\ref{eq_fs}), we obtain
\begin{equation}
\sum_{n}\left( H^{(n-m)}-n\omega \delta _{nm}\right) |\phi _{j}^{(n)}\rangle
=\varepsilon _{j}|\phi _{j}^{(m)}\rangle .
\end{equation}%
The matrix elements of the Floquet Hamiltonian are thus
\begin{equation}
H_{nm}^{(F)}=H^{(n-m)}-n\omega \delta _{nm}.
\end{equation}

In numerical simulations we truncate to photon sectors $|m|\leq M$, giving
\begin{equation}
H_{M}^{(F)}=%
\begin{pmatrix}
H^{(0)}+M\omega & H^{(-1)} & \cdots & H^{(-2M)} \\
H^{(1)} & H^{(0)}+(M-1)\omega & \cdots & H^{(-2M+1)} \\
\vdots & \vdots & \ddots & \vdots \\
H^{(2M)} & H^{(2M-1)} & \cdots & H^{(0)}-M\omega%
\end{pmatrix}%
.  \label{H_MF}
\end{equation}%
Only a few Fourier components are nonzero,
\begin{equation}
H^{(n)}=\left\{
\begin{array}{ll}
\sum_{j=-s}^{s-1}\mu _{j}|j+1\rangle \langle j|, & n=+1, \\
\sum_{j=-s}^{s-1}\mu _{j}|j\rangle \langle j+1|, & n=-1, \\
-\kappa \sum_{j=-s}^{s-1}(|j\rangle \langle j+1|+\mathrm{H.c.}), & n=0, \\
0, & \text{otherwise}.%
\end{array}%
\right.  \label{H_n}
\end{equation}

This formulation provides a clear physical interpretation: the diagonal
blocks $H^{(0)}+n\omega$ correspond to static Hamiltonians shifted by $n$
drive quanta, while the off-diagonal blocks $H^{(\pm 1)}$ describe processes
that change the photon number by $\pm 1$. Therefore the Floquet spectrum
reveals multi-photon resonances whenever $n\omega$ matches gaps of the
static spectrum. In the high-frequency limit, the bands decouple,
reproducing the Floquet-Magnus expansion; in the moderate- and low-frequency
regimes, strong hybridization between photon sectors encodes the essential
physics studied in this work.

\section{The first Floquet-Sambe bottleneck}

\label{bottleneck}

The first Floquet-Sambe bottleneck appears at the weakest spin-assisted
hopping near the boundary, the corresponding scale relation is a scaling
criterion for the frequency $\omega $ at which the boundary hopping first
crosses over from strong Floquet mixing to weak Floquet mixing, while the
bulk still remains strongly mixed. This spatial mismatch can produce
reflection, interference, and compression of the wavefunctions in real
space. Based on the above original Hamiltonian Eq. (\ref{H_1}) and
spin-assisted hopping Eq. (\ref{u_j}). The key feature of the Hamiltonian is
that $\mu _{j}$ is not uniform. It is smallest at the two boundaries and
largest near the center of the chain.

To see this explicitly, we can rewrite
\begin{equation}
j=-s+r,
\end{equation}%
where $r=0,1,\dots ,2s-1$, then we have
\begin{equation}
\mu _{-s+r}^{2}=(r+1)(2s-r).
\end{equation}%
At the left boundary ($r=0$) and the right boundary ($r=2s-1$), the
spin-assisted hoppings satisfy
\begin{equation}
\mu _{-s}=\mu _{s-1}=\sqrt{2s},
\end{equation}%
while $\mu _{j}$ is of order $s$ near the middle of the chain. Thus the
chain contains a wide separation between the weakest boundary hopping and
the strongest bulk hopping
\begin{equation}
\mu _{\text{edge}}\thicksim \sqrt{2s},\text{ }\mu _{\text{bulk}}\thicksim s.
\end{equation}%
For a finite Floquet chain characterized by the parameter $s$, a significant
separation of energy scales emerges. This separation of scales leads to the
existence of an optimal frequency $\omega $ at which the system exhibits
pronounced collective localization features (see Sec. \ref{SL Regime}).

We then back to the time-dependent part of the Hamiltonian Eq. (\ref{H_1}),
which has the Fourier structure
\begin{equation}
H(t)=H^{(0)}+e^{-i\omega t}H^{(-1)}+e^{i\omega t}H^{(+1)}.  \label{H_t}
\end{equation}%
In the Floquet-Sambe basis $|j,m\rangle $, where $m$ is the Floquet index,
different Floquet replicas are separated by the energy
\begin{equation}
\Delta E_{\text{Sambe}}=\omega .
\end{equation}%
The Fourier components $H^{(\pm 1)}$ connect neighboring Floquet replicas.
In particular, the spin assisted hopping $\mu _{j}$ produces transitions
that change the Floquet index by one. Therefore the local strength of
Floquet mixing is controlled by the dimensionless ratio
\begin{equation}
\eta _{j}=\frac{\mu _{j}}{\omega }.
\end{equation}%
If $\eta _{j}\gg 1$, the local Floquet-Sambe channels are strongly mixed. If
$\eta _{j}\ll 1$, the local coupling between neighboring Floquet replicas is
weak and can be treated perturbatively in a high frequency expansion. The
crossover occurs when
\begin{equation}
\eta _{j}\thicksim 1,\text{ }\mu _{j}\thicksim \omega .
\end{equation}%
Since $\mu _{j}$ is smallest at the boundary, the first local crossover
occurs when
\begin{equation}
\omega _{\text{peak}}\thicksim \mu _{-s}=\sqrt{2s},
\end{equation}%
i.e., the first Floquet-Sambe bottleneck. It is the first bond that ceases
to behave as a strongly mixed Floquet channel when $\omega $ is increased.
It should be emphasized that the emergence of this collective localization
feature is controlled not by the largest hopping in the bulk, but by the
weakest hopping near the boundary, since a wavefunction attempting to extend
across the entire finite chain has to pass through the weakest bond.

\subsection{Low-frequency regime}

\label{sec:static}

\begin{figure*}[tbp]
\centering
\includegraphics[bb=60 31 2250 541, width=18cm, clip]{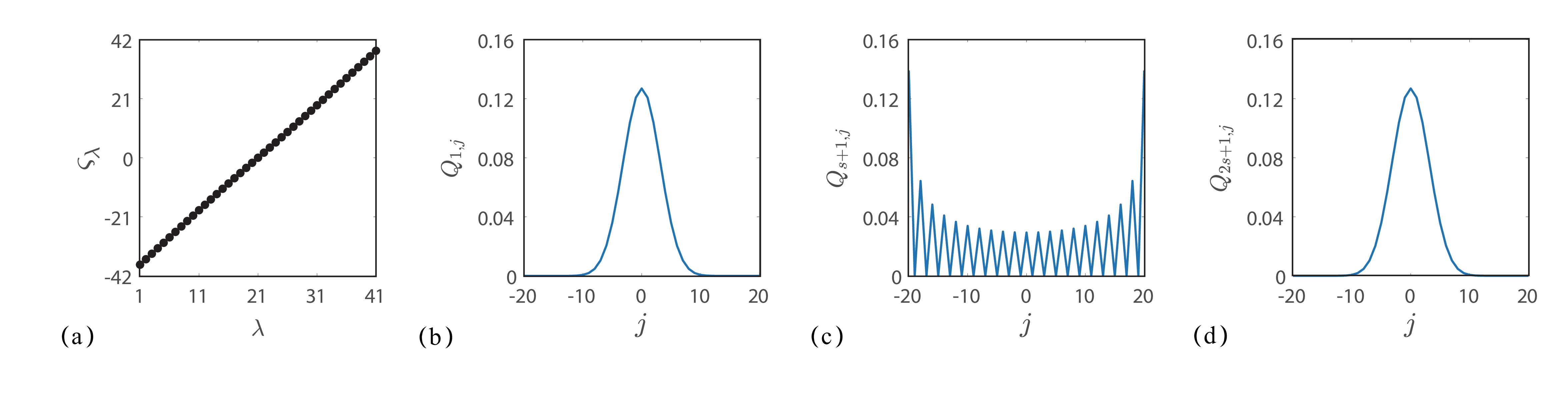}
\caption{Eigenspectrum of the static system ($\protect\omega =0$) and
representative eigenstates. (a) Eigenvalues sorted in ascending order.
(b)-(d) Spatial profiles of three selected eigenstates: the ground state ($%
\protect\lambda =1$), the middle state ($\protect\lambda =s+1$), and the
highest state ($\protect\lambda =2s+1$). These static-limit results
establish the spectral baseline for comparison with the periodically driven
case. Parameters: $\protect\kappa =1$, $s=20$, and $Q_{\protect\lambda %
,j}=|q_{\protect\lambda ,j}|^{2}$.}
\label{fig1}
\end{figure*}

In the absence of periodic driving ($\omega =0$), Eq. (\ref{H_og}) reduces
to the static tight-binding Hamiltonian
\begin{equation}
H(t)=H_{\mathrm{tb}}=\sum_{j=-s}^{s-1}(-\kappa +\mu _{j})(|j\rangle \langle
j+1|)+\mathrm{H.c.},
\end{equation}%
which combines a uniform hopping amplitude $\kappa $ with a spatially
varying hopping strength $\mu _{j}$. The eigenvalues of $H_{\mathrm{tb}}$
are $\varsigma _{\lambda }$ with eigenstates%
\begin{equation}
\left\vert \varphi _{\lambda }\right\rangle =\sum_{j=-s}^{s}q_{\lambda
,j}|j\rangle .
\end{equation}%
In Fig. \ref{fig1}, we present the eigenspectrum of the system in the static
limit ($\omega =0$), along with the eigenstates corresponding to three
representative eigenvalues (corresponding ground state, middle state, and
highest state) selected after sorting the spectrum in ascending order: the $1
$st eigenstate, the$\,(s+1)$-th eigenstate, and the $(2s+1)$-th eigenstate.
These observations establish the baseline spectral features in the static
limit and set the stage for understanding how periodic driving qualitatively
reshapes the energy landscape.

Numerically diagonalizing $H_{\mathrm{tb}}$ reveals a remarkable fact: all
eigenvalues appear to be (nearly) equally spaced, forming a ladder-like
spectrum in the static limit. This occurs because the position-dependent
hopping $\mu _{j}$ realizes a discrete analogue of angular momentum
operators. The corresponding eigenstates interpolate smoothly from strongly
localized modes (dominated by the decreasing hopping $\mu _{j}$ near the
chain edges) to fully extended states near the center. Notabaly, the
eigenstate exhibit a clear nodal ordering similar to the \textquotedblleft
node theorem\textquotedblright\ in Sturm-Liouville problems: the $n$-th
excited state possesses $n$ nodes and becomes progressively more
delocalized. This monotonic change in spatial profile reflects the
competition between the uniform hopping $\kappa $ and the inhomogeneous
hopping $\mu _{j}$, which grows toward the center of the chain. Changing $%
(s,\kappa )$ modifies both the bandwidth and the localization properties.
Increasing $s$ enlarges the effective chain length, broadens the total
bandwidth, and increases the number of nearly equally spaced levels.
Increasing $\kappa $ enhances the uniform hopping and pushes the system
toward extended Bloch-like eigenstates. When $\mu _{j}\gg \kappa $ near the
center, the spectrum becomes more harmonic-like; conversely, the system
approaches a uniform tight-binding chain when $\kappa $ dominates.

Considering the low-frequency condition
\begin{equation}
\omega \ll \mu _{-s},
\end{equation}%
we then have $\mu _{j}/\omega \gg 1$ for all bonds, including the boundary
bonds. The whole system is in a strongly mixed Floquet regime. There is no
special boundary bottleneck, because even the weakest bond is still strongly
mixed, so the spatial compression is not strongest.

It is useful to first ignore the uniform hopping term proportional to $%
\kappa $ and focus on the spin-assisted part. Using the linear
transformation of spin operators
\begin{equation}
S_{x}=\frac{S_{+}+S_{-}}{2},\text{ }S_{y}=\frac{S_{+}-S_{-}}{2i},
\end{equation}%
one can obtain the rotating spin field in the $x-y$ plane, i.e.,
\begin{equation}
H_{s}(t)=2S_{x}\cos (\omega t)-2S_{y}\sin (\omega t).
\end{equation}%
The transformed Hamiltonian is
\begin{equation}
H_{\text{rot}}=R^{\dagger }H_{s}(t)R-iR^{\dagger }\dot{R},  \label{transf_H}
\end{equation}%
with the rotating-frame transformation $R(t)=e^{i\omega tS_{z}}$. We then
use the conditions
\begin{equation}
R^{\dagger }S_{+}R=e^{-i\omega t}S_{+},\text{ }R^{\dagger }S_{-}R=e^{i\omega
t}S_{-},
\end{equation}%
and
\begin{equation}
R^{\dagger }H_{s}(t)R=S_{+}+S_{-}=2S_{x},\text{ }-iR^{\dagger }\dot{R}%
=\omega S_{z}.
\end{equation}%
Eq. (\ref{transf_H}) can be rewritten as
\begin{equation}
H_{\text{rot}}=2S_{x}+\omega S_{z},
\end{equation}%
and the eigenstates of $H_{\text{rot}}$ are spin states aligned with the
effective field
\begin{equation}
\mathbf{B}_{\text{eff}}=(2,0,\omega ),
\end{equation}%
where $\tan \theta =2/\omega $, the eigenstates can be written as
\begin{equation}
|\ell ,\theta \rangle =e^{-i\theta S_{y}}|\ell \rangle _{z},
\end{equation}%
with $\ell =-s,-s+1,...,s$. In the site basis $\{|j\rangle \}$ (the $S_{z}$
basis), the expansion coefficients are Wigner $d$-matrix elements%
\begin{equation}
\langle j|\ell ,\theta \rangle =d_{j\ell }^{s}(\theta ).
\end{equation}%
In the very low-frequency limit ($\theta \approx \pi /2$), the eigenstates
are close to $S_{x}$ eigenstates, which are broadly distributed in the $S_{z}
$ basis. Thus the low-frequency states are not strongly localized. The
uniform hopping term $H^{(0)}=-\kappa (C_{+}+C_{-})$\ with
\begin{equation}
C_{+}=\sum_{j=-s}^{s-1}|j\rangle \langle j+1|,\text{ }C_{-}=%
\sum_{j=-s}^{s-1}|j+1\rangle \langle j|
\end{equation}%
modifies this ideal rotating-spin picture.

\begin{figure*}[tbp]
\centering
\includegraphics[bb=46 53 1980 1107, width=19 cm, clip]{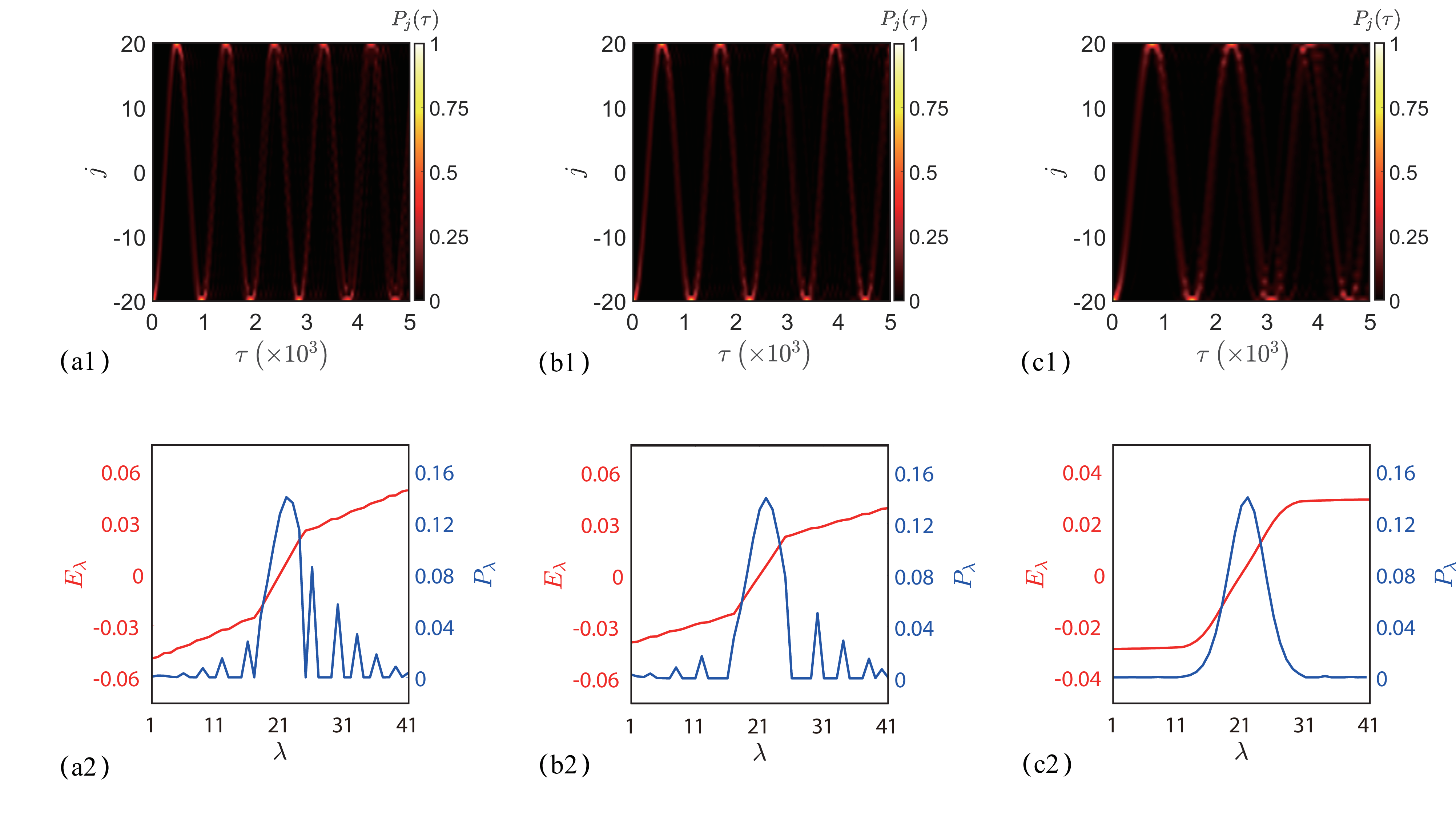}
\caption{{}Frequency dependence of residual \textrm{PST} character in the
low-frequency regime. Each column corresponds to a typical driving
frequency: $\protect\omega =0.1$ [panels (a1-a2)], $\protect\omega =0.08$
[panels (b1-b2)], and $\protect\omega =0.07$ [panels (c1-c2)]. Top row:
Space-time profile $P_{j}(\protect\tau )=|\langle j|\Psi (\protect\tau %
)\rangle |^{2}$ showing the propagation of an initially boundary-localized
state (site state localized at $j=-20$). As $\protect\omega $ decreases, the
\textrm{PST} signature becomes increasingly pronounced: the peak probability
at the opposite end rises systematically from $0.77$\ at $\protect\omega =0.1
$ to $0.83$ at $\protect\omega =0.08$ and approaches $0.84$ at $\protect%
\omega =0.07$, signaling the gradual recovery of near-perfect state
transfer. Bottom row: Instantaneous eigenvalues $E_{\protect\lambda }$ at $%
\protect\tau =0$ (red) and the overlap $P_{\protect\lambda }=|\langle
\protect\phi _{\protect\lambda }|\Psi (0)\rangle |^{2}$ of the initial state
(blue). The initial state populates exclusively the central band of
eigenstates with approximately equal energy spacing, which governs the
periodic revival dynamics. With decreasing $\protect\omega $, the spectral
distribution of the initial state narrows progressively, converging toward
the static $S_{x}$ system. Other parameters: $\protect\kappa =1$ and $s=20$.}
\label{fig2}
\end{figure*}

The quest for perfect state transfer (\textrm{PST}) has evolved from
Hermitian chains with engineered couplings \cite{Christdl} to $\mathcal{PT}$%
-symmetric non-Hermitian networks where \textrm{PST} persists conditionally
in the unbroken phase \cite{ZXZ}. In low-frequency regime, the Hamiltonian
Eq. (\ref{H_1}) reduces exactly to the Christandl model. A natural question
arises as to whether the \textrm{PST} character survives under small but
finite driving frequencies $\omega $. To address this, we examine the three
typical low-frequency regime Floquet frequencies $\omega =0.07$, $0.08$, and
$0.1$. The numerical results demonstrate that the system exhibits a
pronounced revival of the \textrm{PST} signature: an initial state $|\Psi
(0)\rangle $ localized at one boundary evolves into a translated copy at the
opposite boundary at times
\begin{equation}
\tau \approx \frac{(2n+1)\pi }{\Delta E},
\end{equation}%
where $\tau $ is the \textrm{PST} time of the static Christandl chain,
realizing near-perfect state transfer. We then define the time evolution
state%
\begin{equation}
|\Psi (t)\rangle =\sum_{\lambda }c_{\lambda }e^{-i\varepsilon _{\lambda
}\tau }|\phi _{\lambda }\rangle .
\end{equation}%
The residual \textrm{PST} is illustrated in Fig. \ref{fig2} through time
evolution, demonstrating that the Floquet drive acts as a weak perturbation
to the underlying \textrm{PST} mechanism at low-frequency regime, with the
transfer fidelity remaining close to unity despite the breaking of exact
integrability.

\subsection{High-frequency regime}

\label{sec:highfreq}

\begin{figure*}[tbp]
\centering
\includegraphics[bb=65 32 1698 527, width=18 cm, clip]{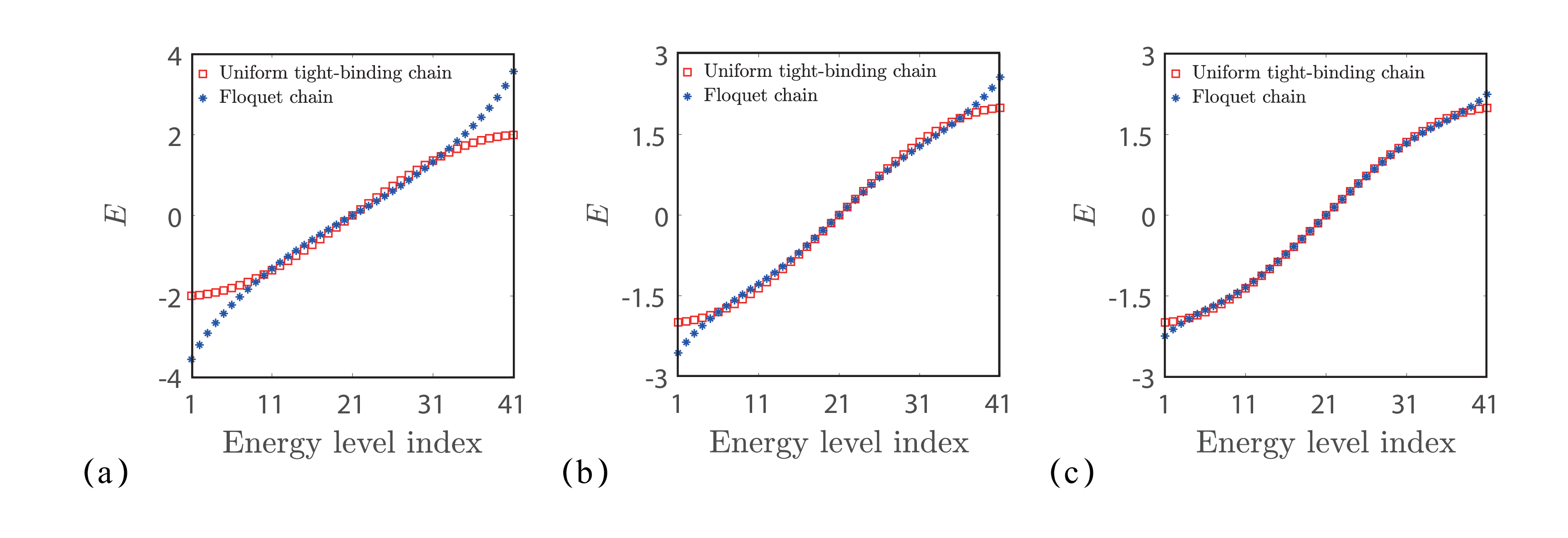}
\caption{Floquet quasi-spectrum (blue star) compared with the effective
static tight-binding spectrum governed by $H_{\mathrm{eff},1}$ (red square)
for driving frequencies (a) $\protect\omega =20$, (b) $\protect\omega =50$,
and (c) $\protect\omega =100$. With increasing $\protect\omega $, the
Floquet spectrum converges to the cosine band of the effective uniform
tight-binding chain, consistent with the high-frequency expansion. Other
parameters: $\protect\kappa =1$ and $s=20$.}
\label{fig3}
\end{figure*}

In the opposite regime, the Hamiltonian under very fast driving varies more
rapidly than the intrinsic timescales of the chain. The long-time dynamics
is then captured by the Floquet-Magnus expansion
\begin{equation}
H_{F}=H_{\mathrm{eff}}^{(1)}+\frac{1}{\omega }H_{\mathrm{eff}}^{(2)}+\frac{1%
}{\omega ^{2}}H_{\mathrm{eff}}^{(3)}+\cdots ,  \label{magn_exp}
\end{equation}%
where
\begin{equation}
\begin{cases}
H_{\mathrm{eff}}^{(1)}=H_{0}, \\[4pt]
H_{\mathrm{eff}}^{(2)}=\frac{1}{2}\sum_{n\neq 0}\frac{[H^{(n)},H^{(-n)}]}{n},
\\[4pt]
H_{\mathrm{eff}}^{(3)}=\frac{1}{3}\sum_{n,m\neq 0}\frac{%
[H^{(m)},[H^{(n-m)},H^{(-n)}]]}{mn} \\
+\frac{1}{2}\sum_{n,m\neq 0}\frac{[[H^{(m)},H^{(n)}],H^{(-n-m)}]}{m(n+m)}.%
\end{cases}%
\end{equation}%
Considering the spin-assisted hopping
\begin{equation}
H^{(+1)}=S_{+},\text{ }H^{(-1)}=S_{-},
\end{equation}%
and angular-momentum commutator
\begin{equation}
\lbrack S_{+},S_{-}]=2S_{z},
\end{equation}%
we can obtain
\begin{equation}
H_{F}=H_{\mathrm{eff},1}\approx -\kappa (C_{+}+C_{-})-\frac{2}{\omega }S_{z}+%
\mathcal{O}(\omega ^{-2}).
\end{equation}%
It is easy to check that the effective Hamiltonian $H_{\mathrm{eff},1}$
approaches a nearly uniform tight-binding chain
\begin{equation}
H_{\mathrm{eff},1}\approx -\kappa (C_{+}+C_{-}),
\end{equation}%
in the limit $\omega \rightarrow \infty $. Consequently, the eigenstates
become extended Bloch waves with dispersion
\begin{equation}
E(k)=-2\kappa \cos k.
\end{equation}%
In Fig. \ref{fig3}, we show the Floquet quasi-energy spectra for three
typical driving frequencies ($\omega =20$, $50$, and $100$). The
quasi-energies (blue star) are compared with the spectrum of the uniform
tight-binding chain governed by $H_{\mathrm{eff},1}$ (red square). With
increasing $\omega $, the two spectra progressively approach each other.
These numerical findings indicate that, in the high-frequency limit, the
Floquet quasi-spectrum reduces to a cosine band of a tight-binding model.

The two limits $\omega \rightarrow 0$ and $\omega \rightarrow \infty $
define qualitatively distinct dynamical regimes: For static/slow driving%
\textbf{:} instantaneous eigenstates of $H(0)$ are followed adiabatically,
the spectrum features harmonic-like level spacings and eigenstates
interpolate from localized edge modes to extended bulk modes. For
high-frequency regime: the drive averages out, the system behaves as a
uniform tight-binding model with extended Bloch waves and cosine dispersion.

At intermediate frequencies, the primary regime explored in this work, the
periodic drive is fast enough to produce well-defined Floquet sectors yet
slow enough to allow resonant hybridization between them. This competition
leads to the emergence of Floquet quasi-energy ladders with nearly equal
spacings. These ladders underpin the Bloch-oscillation-like dynamics
presented in Sec. \ref{SL Regime}.

\section{Emerged Stark-Ladder Regime}

\label{SL Regime}

Periodic driving introduces a competition among three key ingredients of the
lattice model: the uniform hopping $\kappa $, the monotonic hopping
deformation $\mu _{j}$, and the photon-sector hybridization encoded in the
Floquet modes. Within specific finite-frequency windows, this competition
gives rise to an emergent quasi-energy structure that closely resembles a
Wannier-Stark ladder, featuring nearly equally spaced levels together with
robust localization and coherent transport phenomena.

\subsection{Effective Hamiltonian and Emergent Linear Potential}

For a finite driving frequency $\omega $, the periodic potential $H_{s}(t)$
oscillates rapidly, such that its time-averaged effect can be seen as
perturbative. Retaining the first two terms of the Floquet-Magnus expansion
in Eq. (\ref{magn_exp}), the effective Hamiltonian reads
\begin{equation}
H_{\mathrm{eff},2}\approx H_{0}+\frac{1}{\omega }\sum_{n\neq 0}\frac{%
[H_{n},H_{-n}]}{2},  \label{Heff2}
\end{equation}%
where contributions of order $\mathcal{O}(\omega ^{-2})$ and higher are
neglected.

Substituting the Fourier components $H^{(n)}$ of Eq. (\ref{H_n}) into Eq. (%
\ref{Heff2}) yields a time-independent lattice Hamiltonian
\begin{equation}
H_{\mathrm{eff},2}=-\kappa _{\mathrm{eff}}\sum_{j=-s}^{s-1}[\left( |j\rangle
\langle j+1|)+\mathrm{H.c.}\right) ]+F_{\mathrm{eff}}\sum_{j=-s}^{s}j|j%
\rangle \langle j|,  \label{H_li}
\end{equation}%
with renormalized parameters
\begin{equation}
\kappa _{\mathrm{eff}}=\kappa ,\text{ }F_{\mathrm{eff}}=\frac{2}{\omega }.
\end{equation}%
Thus, to order $\mathcal{O}(\omega ^{-1})$ the driven system maps exactly
onto a discrete Stark ladder: uniform nearest-neighbor hopping $\kappa _{%
\mathrm{eff}}$ subjected to a linear potential of strength $F_{\mathrm{eff}}$%
\textrm{. }Solving the eigenvalue problem
\begin{equation}
H_{\mathrm{eff},2}|\Phi _{\lambda }\rangle =E_{\lambda }|\Phi _{\lambda
}\rangle ,  \label{E_eig}
\end{equation}%
and expanding $|\Phi _{\lambda }\rangle =\sum_{j}c_{j,\lambda }|j\rangle $
gives the recurrence relation
\begin{equation}
-\kappa _{\mathrm{eff}}(c_{j+1,\lambda }+c_{j-1,\lambda })+jF_{\mathrm{eff}%
}c_{j,\lambda }=E_{\lambda }c_{j,\lambda }.  \label{rec1}
\end{equation}%
Comparing Eq. (\ref{rec1}) with the Bessel-function identity
\begin{equation}
J_{\nu +1}(x)+J_{\nu -1}(x)=\frac{2\nu }{x}J_{\nu }(x),  \label{Bes1}
\end{equation}%
yields the exact solutions
\begin{equation}
E_{\lambda }=\lambda F_{\mathrm{eff}}  \label{E_eq}
\end{equation}%
and
\begin{equation}
|\Phi _{\lambda }\rangle =\sum_{j=-s}^{s}J_{j-\lambda }(-\frac{2\kappa _{%
\mathrm{eff}}}{F_{\mathrm{eff}}})|j\rangle .  \label{Phi_eq}
\end{equation}%
It is easy to check that the eigenstates satisfy the translational
covariance
\begin{equation}
c_{j+\eta ,\lambda }=c_{j,\lambda -\eta },  \label{trs}
\end{equation}%
showing that eigenstates separated by $\eta F_{\mathrm{eff}}$ are exactly
related by $\eta $-site translations. Finite-size edges introduce only weak
corrections and preserve the Stark-ladder structure over wide parameter
ranges.

Bloch oscillations (\textrm{BO}), originally associated with particle motion
in a uniform field \cite{Zener}, acquire a dynamical counterpart in
periodically driven systems. Here, Bloch-like oscillations emerge not from
an external potential gradient but from resonant hybridization among photon
replicas of the static eigenstates.

In Sambe space, the Floquet Hamiltonian reads
\begin{equation}
H_{nm}^{(F)}=H^{(n-m)}-n\omega \delta _{nm},
\end{equation}%
where $H^{(n-m)}$ are the Fourier components and $n,m$ denote photon
sectors. When a cluster of static eigenstates satisfies the resonance
condition
\begin{equation}
\mathcal{E}_{\alpha }^{(0)}-\mathcal{E}_{\beta }^{(0)}\approx n\omega ,
\end{equation}%
their photon replicas align and hybridize, forming a quasi-energy manifold
with nearly uniform spacing
\begin{equation}
E_{\lambda }\approx \lambda \Delta E+\varepsilon _{0},
\end{equation}%
where $\varepsilon _{0}$ is the constant energy offset of the spectrum. The
spacing $\Delta E=F_{\mathrm{eff}}$ serves as an effective Bloch-like
frequency arising purely from virtual-photon processes and finite-size
boundary conditions. The dynamical manifestations depend on two properties:
(i)\ the site-shift symmetry $c_{j+1,\lambda }\approx c_{j,\lambda -1}$
inherited from Eq. (\ref{trs}), which guarantees that ladder states are
translated copies; (ii)\ strong spectral overlap of the initial state with
the ladder manifold.

\begin{figure*}[tbp]
\centering
\includegraphics[bb=73 50 1241 538, width=16 cm, clip]{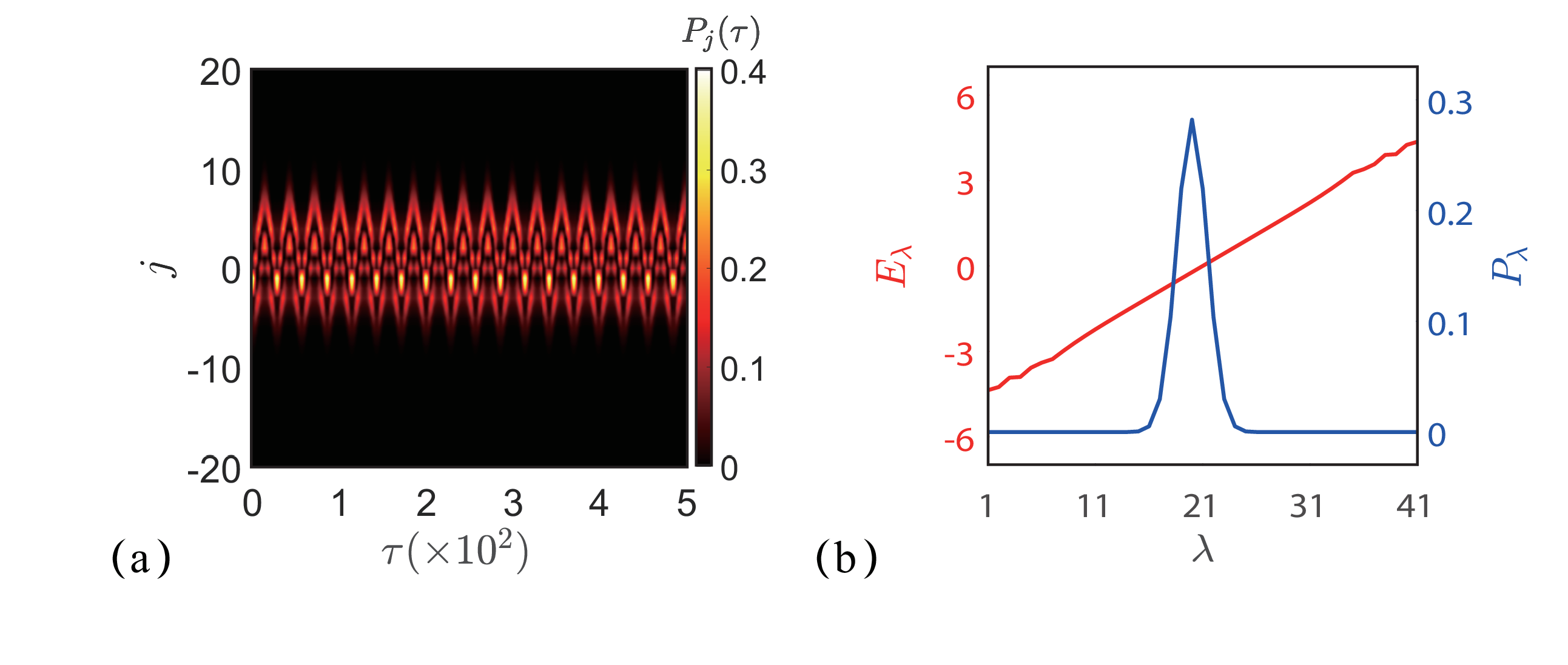}
\caption{Bloch-like oscillation dynamics and spectral structure at a
specific driving frequency. (a) Space-time profile $P_{j}(\protect\tau %
)=|\langle j|\Psi (\protect\tau )\rangle |^{2}$, showing the propagation of
an initial state along the spin chain. The wavepacket exhibits pronounced
Bloch-like oscillation, with the probability density oscillating over
multiple cycles. (b) Instantaneous eigenvalues $E_{\protect\lambda }$ at $%
\protect\tau =0$ (red) and the overlap $P_{\protect\lambda }=|\langle
\protect\phi _{\protect\lambda }|\Psi (0)\rangle |^{2}$ of the initial state
with each eigenstate (blue). The initial state populates exclusively a
narrow band of central eigenstates with approximately equal energy spacing,
which governs the coherent revival dynamics observed in panel (a).
Parameters: $\protect\omega =9$, $\protect\kappa =1$, and $s=20$.}
\label{fig4}
\end{figure*}

The time evolution state
\begin{equation}
|\Psi (t)\rangle =\sum_{\lambda }c_{\lambda }e^{-i\varepsilon _{\lambda
}t}|\phi _{\lambda }\rangle
\end{equation}%
is then dominated by the phase factor $e^{-i\lambda \Delta Et}$. Under an
effective field $F_{\mathrm{eff}}$, the initial state undergoes periodic
revivals with period $T_{\mathrm{rev}}=2\pi /\Delta E$. As illustrated in
Fig. \ref{fig4}, rather than transferring to the opposite end, the
wavepacket undergoes sustained Bloch-like oscillations: the probability
density oscillates periodically across the chain without accumulating at the
far boundary, exhibiting a recurrent revival pattern over multiple driving
cycles.

\subsection{MIPR Diagnostic and Green-Function IPR Diagnostic}

In the high-frequency limit $\omega \rightarrow \infty $, the drive averages
to zero, recovering $H_{0}$ with extended Bloch eigenstates. At finite $%
\omega $, however, the induced Stark field $F_{\mathrm{eff}}$ localizes
eigenstates.

\begin{figure}[tbp]
\centering
\includegraphics[bb=40 73 560 534, width=7.5 cm, clip]{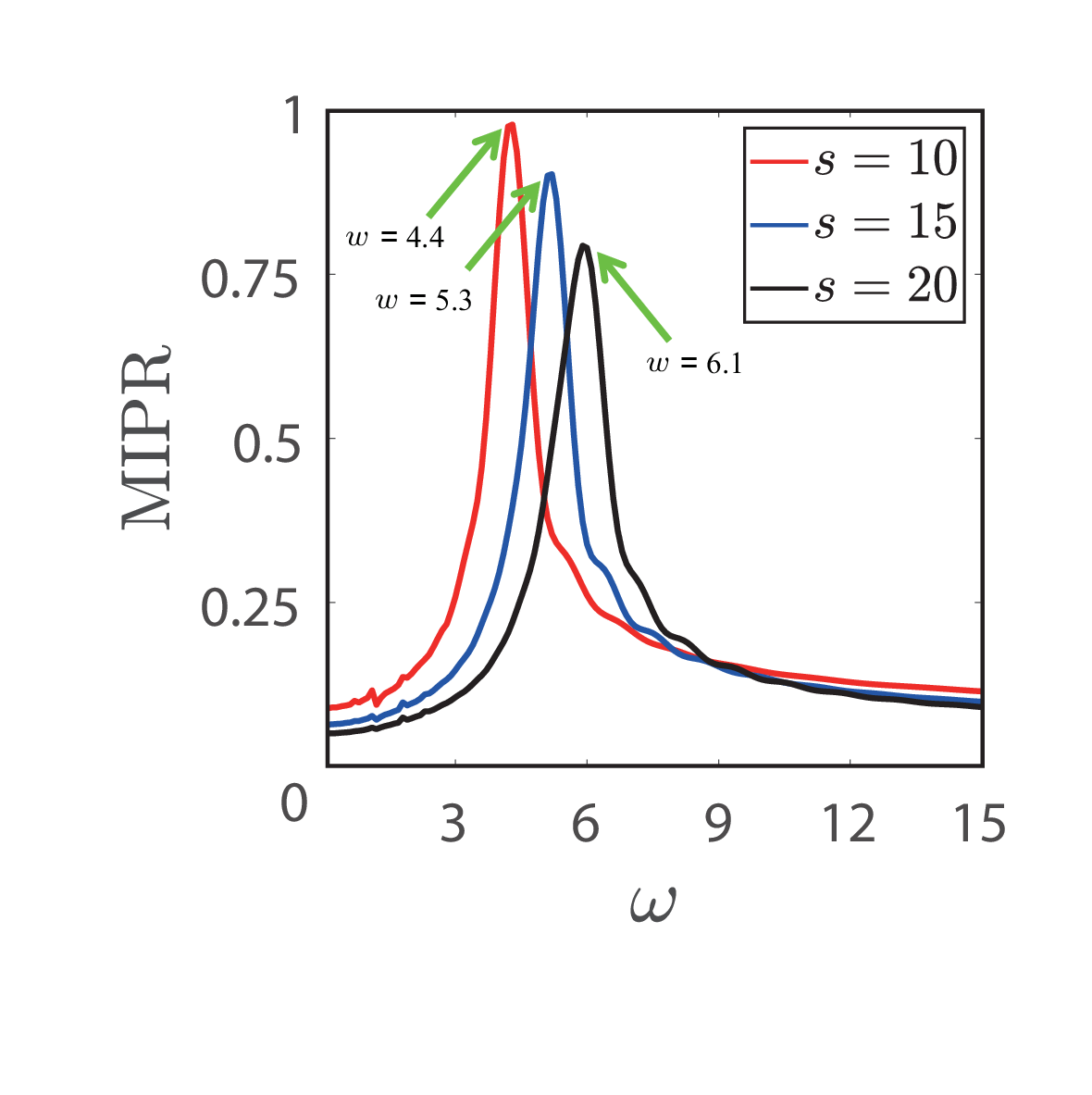}
\caption{\textrm{MIPR} as a function of driving frequency $\protect\omega $
for spin quantum numbers $s=10$ (red), $s=15$ (blue), and $s=20$ (black).
The green arrow denotes the value of $\protect\omega \approx 4.4$, $5.3$,
and $6.1$, respectively. Each curve exhibits a nonmonotonic profile with a
well-defined peak at an intermediate frequency $\protect\omega _{\text{peak}}
$, separating the low-frequency regime dominated by high-order Magnus
corrections from the high-frequency regime where Stark localization weakens.
The peak position shifts to larger $\protect\omega $ with increasing $s$.
The numerical results are in excellent agreement with the first
Floquet-Sambe bottleneck condition $\protect\omega _{\text{peak}}\thicksim
\protect\sqrt{2s}$. Other parameters: $\protect\kappa =1$ and $s=20$.}
\label{fig5}
\end{figure}

On the one hand, we can diagnose localization using the closed-system
\textrm{IPR}
\begin{equation}
\mathrm{IPR}(j)=\frac{\sum_{l}|\langle \phi _{j}|l\rangle |^{4}}{%
(\sum_{l}|\langle \phi _{j}|l\rangle |^{2})^{2}},
\end{equation}%
and the \textrm{MIPR}
\begin{equation}
\mathrm{MIPR}=\frac{\sum_{j=-s}^{s}\mathrm{IPR}(j)}{2s+1},
\end{equation}%
which captures global localization trends. In Fig. \ref{fig5}, we present
the \textrm{MIPR} as a function of the Floquet driving frequency $\omega $
for three typical values of the spin quantum numbers $s=10$, $15$, and $20$.
We note that the numerical \textrm{MIPR} exhibits three characteristic
regimes: (1) Low-frequency regime ($\omega <\omega _{\text{peak}}$):
higher-order Magnus terms ($\propto 1/\omega ^{n}$) distort the Stark
structure, reducing localization and thus lowering \textrm{MIPR}.\ (2)
High-frequency regime ($\omega >\omega _{\text{peak}}$): as $\omega $
increases, $F_{\mathrm{eff}}=2/\omega $ weakens, reducing localization; once
the Stark localization length exceeds the system size, finite-size effects
broaden the states, lowering \textrm{MIPR}. (3) Critical regime ($\omega
=\omega _{\text{peak}}$): a balance between Stark localization and
higher-order corrections yields maximal localization and a pronounced
\textrm{MIPR} peak, marking the optimal frequency window for emergent
Stark-ladder. This optimal frequency window is in excellent agreement with
the first Floquet-Sambe bottleneck condition derived in Sec. \ref{bottleneck}%
.

To connect the closed-system localization to scattering probes, we further
formulate an open-system Floquet-Sambe Green-function diagnostic. We
construct a spatial density from the retarded Green function generated by an
injection from the left lead. The corresponding Green-function \textrm{IPR}
measures the spatial concentration of the injected Floquet scattering state
inside the driven region. This quantity is not identical to the
closed-system \textrm{MIPR}, because it is filtered by the lead bandwidth,
the boundary self-energy, and the injection profile. Nevertheless, it
displays the same nonmonotonic localization trend. Its peak is shifted to
larger driving frequencies because the relevant virtual sideband denominator
is set by the distance from $E\pm \omega $ to the static band and the lead
self-energy, rather than by $\omega $ alone.

\begin{figure}[tbp]
\centering
\includegraphics[bb=236 81 759 534, width=7.5 cm, clip]{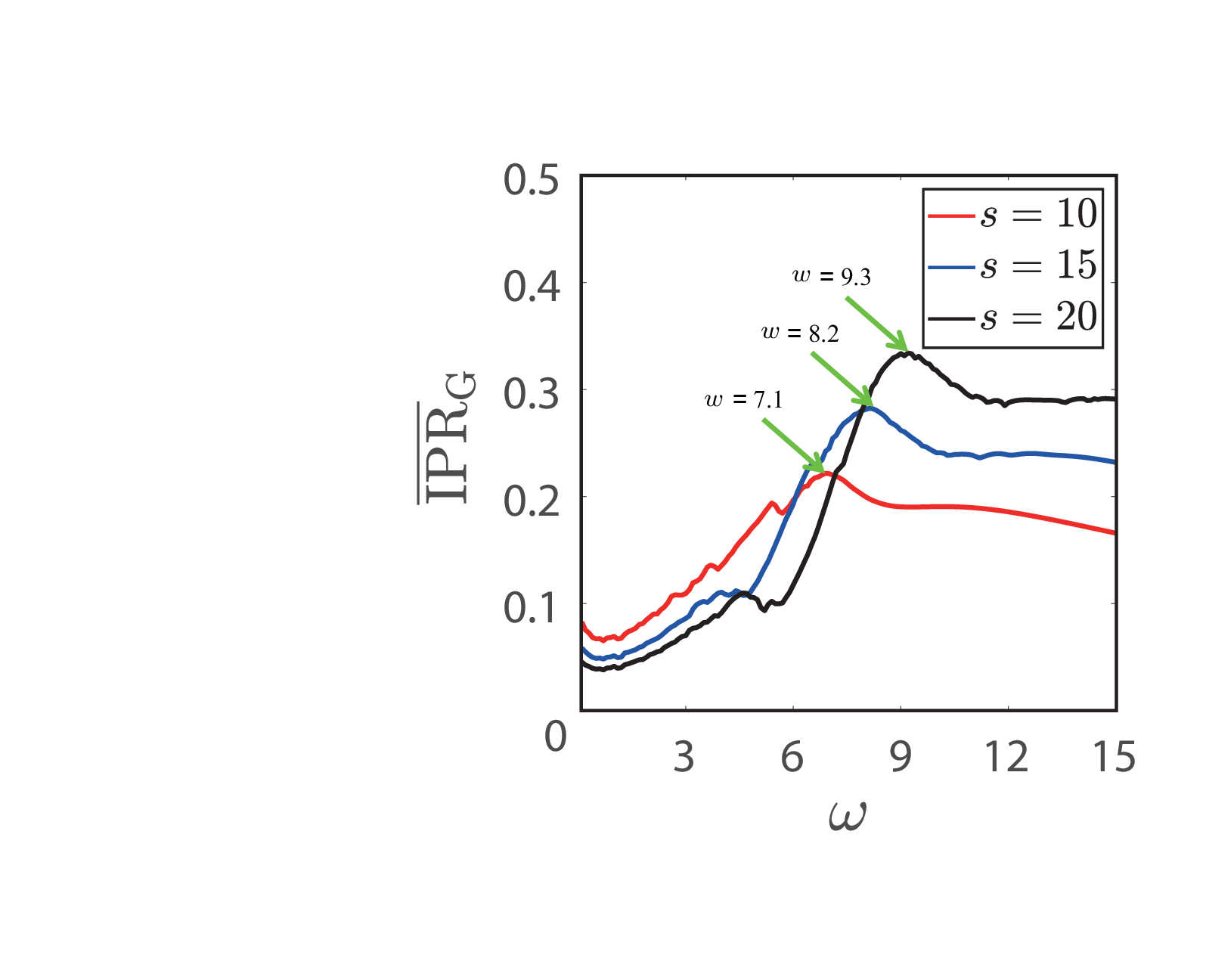}
\caption{Energy-integrated \textrm{IPR} [$\overline{\text{\textrm{IPR}}}_{G}(%
\protect\omega )]$, as a function of driving frequency $\protect\omega $ for
three spin quantum numbers $s=10$ (red), $s=15$ (blue), and $s=20$ (black),
the green arrow denotes the position of $\protect\omega _{c}\approx 7.1$, $%
8.2$, and $9.3$, respectively. The energy range $E\in \lbrack -2,2]$ is
discretized using a uniform mesh with $\Delta E=4/150$. The peak position
increase monotonically with $s$, which is consistent with the theoretical
analysis presented in the main text. Other parameters: $\protect\kappa =1$
and $M=50$, and $s=20$.}
\label{fig6}
\end{figure}

For clarity, we can define the injection-resolved spectral density%
\begin{equation}
\rho _{j}(E,\omega )=\Gamma _{L}(E)\sum_{m}|G_{(j,m)(j_{L},0)}^{r}(E,\omega
)|^{2},
\end{equation}%
i.e., the total probability weight accumulated at each lattice site $j$
after electrons injected from the leftmost site $j_{L}$ of the $m=0$ Floquet
channel are scattered by the driven system $H_{M}^{(F)}$, summed over all
Floquet sidebands $m$, where the retarded Green's function in matrix form is
given by
\begin{equation}
G^{r}=[EI-H_{M}^{(F)}-\Sigma _{L}(E)-\Sigma _{R}(E)]^{-1}
\end{equation}%
with the injection energy $E$, the left (right) self-energy $\Sigma
_{L(R)}(E)$, and the coupling functions $\Gamma _{L(R)}(E)=2$Im$[\Sigma
_{L(R)}(E)]$, we define the open-system Green-function \textrm{IPR} to
quantify the spatial extent of the scattered wave
\begin{equation}
\text{\textrm{IPR}}_{G}(E,\omega )=\frac{\sum_{j}\rho _{j}^{2}(E,\omega )}{%
[\sum_{j}\rho _{j}(E,\omega )]^{2}},
\end{equation}%
where the normalization coefficient is denoted by $[\sum_{j}\rho
_{j}(E,\omega )]^{2}$. Further consider the energy-integrated\ \textrm{IPR}
in the real space
\begin{equation}
\overline{\text{\textrm{IPR}}}_{G}(\omega )=\frac{\int_{-2\kappa ^{\prime
}}^{2\kappa ^{\prime }}W(E)\text{\textrm{IPR}}_{G}(E,\omega )dE}{%
\int_{-2\kappa ^{\prime }}^{2\kappa ^{\prime }}W(E)dE},
\end{equation}%
here $W(E)$ is a normalized spectral weight over the propagating band of the
lead, and we set $\int_{-2\kappa ^{\prime }}^{2\kappa ^{\prime }}W(E)dE=1$.
It specifies how different incident energies contribute to the averaged
Green-function \textrm{IPR}. Unless otherwise stated, we use a uniform
weight $W(E)=1/(4\kappa ^{\prime })$, where $\kappa ^{\prime }$ is the
uniform hopping strength of the lead, so that the integral measures the
band-averaged spatial concentration of the Green-function response. Filtered
by the lead bandwidth, boundary self-energy, and injection profile, this
quantity shares the same peak trend as the closed-system \textrm{IPR} but
with a shifted peak
\begin{equation}
\omega _{\text{peak}}^{G}=\sqrt{2s}+O(\kappa ^{\prime }).
\end{equation}%
In Fig. \ref{fig6}, we shows the energy-integrated \textrm{IPR }[$\overline{%
\text{\textrm{IPR}}}_{G}(\omega )$], as a function of the driving frequency $%
\omega $ for three typical spin quantum numbers $s$. All curves exhibit a
pronounced peak at intermediate frequencies $\omega _{c}$, signaling a
frequency window where the scattered wave is most strongly localized in real
space. Notably, the peak position monotonically increase with $s$,
indicating that a larger spin dimension shifts the optimal driving frequency
$\omega _{c}$ to higher values.\ This numerical result is consistent with
the theoretical analysis presented above.
%{Energy-integrated \textrm{IPR} [$\overline{\text{\textrm{IPR}}%
%}_{G}(\omega )]$, as a function of driving frequency $\omega $ for three spin
%quantum numbers $s=10$ (red), $s=15$ (blue), and $s=20$ (black), the
%green arrow denotes the position of $\omega _{c}\approx 7.1$%
%, $8.2$, and $9.3$, respectively. The energy range $E\in \lbrack -2,2]$ is
%discretized using a uniform mesh with $\Delta E=4/150$. The peak position
%increase monotonically with $s$, which is consistent with the theoretical
%analysis presented in the main text. Other parameters: $\kappa =1$ and $M=50$%
%, and $s=20$.}

\section{Summary}

\label{Summary}

In summary, we have investigated a finite Floquet chain in which uniform
nearest-neighbor hopping coexists with a periodically rotating, $\mathrm{SU}%
(2)$-dictated spin-assisted hopping profile $\mu _{j}=\sqrt{s(s+1)-j(j+1)}$.
This spatially inhomogeneous coupling---weakest at the two boundaries and
strongest in the bulk---produces a frequency-dependent Floquet-Sambe
bottleneck that controls the inter-sector hybridization. In the closed
system, the mean inverse participation ratio (\textrm{MIPR}) of the Floquet
eigenstates exhibits a striking nonmonotonic dependence on the driving
frequency $\omega $: the states remain extended in the low-frequency limit ($%
\omega \rightarrow 0$), become maximally localized at an intermediate
frequency $\omega _{\mathrm{peak}}\sim \mu _{-s}=\sqrt{2s}$---a scale set by
the first boundary bottleneck---and recover an extended character in the
high-frequency limit, where the Magnus expansion generates a Stark term $%
S_{z}/\omega $ that weakens as $\omega $ increases. To connect these
spectral properties to measurable transport, we further construct an
open-system Floquet-Sambe Green-function \textrm{IPR} from the spatial
density of the injected scattering state. This diagnostic recovers the same
nonmonotonic trend, with the localization peak shifted to higher frequencies
by the static bandwidth and the lead self-energy, establishing consistency
between the closed- and open-system perspectives. These findings establish
the driven synthetic spin chain as a frequency-tunable platform for coherent
information storage and retrieval: a functionality rooted in the interplay
between Floquet-Sambe virtual channels, boundary-controlled localization,
and frequency-selective transport, and directly realizable in emerging
multi-level superconducting-circuit architectures.

\acknowledgments This work was supported by the National Natural Science
Foundation of China (Grants No. 12305026, 12275193, 11975166, 12505015),
Science \& Technology Development Fund of Tianjin Education Commission for
Higher Education (Grants No. 2024KJ060), and the Guangdong Basic and Applied
Basic Research Foundation (Grants No. 2024A1515110222).

\end{document}